\documentclass[aps,pra,showpacs,amsfonts,amssymb,superscriptaddress,twocolumn]{revtex4}

\usepackage{amsfonts}
\usepackage{amsmath}
\usepackage{bm}
\usepackage{graphics,graphicx}
\newcommand{\be}{\begin{equation}}
\newcommand{\ee}{\end{equation}}
\newcommand{\bea}{\begin{eqnarray}}
\newcommand{\eea}{\end{eqnarray}}

\newcommand{\nn}{\nonumber}

\newcommand{\up}{\uparrow}
\newcommand{\down}{\downarrow}

\newcommand{\ket}[1]{\left\vert #1    \right\rangle }

\begin{document}

\title{Tavis-Cummings model and collective multi-qubit entanglement in trapped ions}

\author{A. Retzker} \email{a.retzker@imperial.ac.uk}

\affiliation{Institute for Mathematical Sciences, Imperial College
London, SW7 2PE, UK} \affiliation{QOLS, The Blackett Laboratory,
Imperial College London, Prince Consort Rd., SW7 2BW, UK}

\author{E. Solano \footnote{Present address: Physics Department,
Ludwig-Maximilian University, Munich, Germany,
enrique.solano@physik.lmu.de}}

\affiliation{Max-Planck-Institut f\"{u}r Quantenoptik,
Hans-Kopfermann-Strasse 1, D-85748 Garching, Germany}

\affiliation{Secci\'on F\'{\i}sica, Departamento de Ciencias,
Pontificia Universidad Cat\'olica del Per\'u, Apartado Postal 1761,
Lima, Peru}

\author{B. Reznik}

\affiliation{Department of Physics and Astronomy, Tel-Aviv
University, Tel Aviv 69978, Israel}

\begin{abstract}
We present a method of generating collective multi-qubit
entanglement via global addressing of an ion chain performing {\it blue} and {\it red} Tavis-Cummings interactions, where several qubits are
coupled to a collective motional mode. We show that a wide family of
Dicke states and irradiant states can be generated by single global
laser pulses, unitarily or helped with suitable postselection
techniques.
\end{abstract}


 \maketitle

\date{\today}

\section{Introduction}

Multi-partite entangled states play an important role in quantum
information. They are useful in various quantum information
applications, such as in Heinsenberg-limited spectroscopy~\cite{leibfried04}, secure communication \cite{wang06},
and various schemes related to "one-way" quantum
computing~\cite{briegel01}. Fresh theoretical developments on the
generation of multipartite entangled states show that sequential
techniques may prove to be general and practical for building
arbitrary multi-qubit states~\cite{schoen05}. For instance, a recent
experiment~\cite{haeffner05} has realized a $W$ state of eight
qubits, encoded in the internal ionic levels, by performing a
sequence of two-qubit gates on different ion pairs. However, given a
set of available interactions in a physical system, there are
particular families of entangled states that could be built globally
and in fewer steps~\cite{leibfried05}. In the context of cavity QED
(CQED), for example, the coupling of a single cavity mode with a
two-level atom, the Jaynes-Cummings model (JC), can be extended to
the $N$ atom case, leading to the Tavis-Cummings model, with
different dynamics and entanglement
features~\cite{tavis68,tessier03}.

In this article, we study methods of generating specific classes of
multi-qubit entangled states in trapped ions with collective
interactions, which are potentially faster and more efficient than
individual techniques. They consist of two key ingredients: firstly,
the use of global rather than individual addressing of ions and,
secondly, the presence of {\it invariant subspaces}, i.e., combined
(vibronic) internal and motional finite subspaces that are {\it
closed} with respect to certain dynamical operations.

In Sec.~\ref{collmaps}, we describe realistic collective vibronic
interactions coupling the internal degrees of freedom of $N$ ions
with a collective motional mode. Specifically, we consider the blue
and red excitation versions of the Tavis-Cummings model, taking
distance from usual predictions in the Dicke model. In
Sec.~\ref{invsub}, we study the invariant subspaces, associated with
the proposed interactions, in the search of classes of multipartite
entangled states that may be efficiently generated. It will turn out
that one of them is the family of symmetric Dicke
states~\cite{dicke54,stockton03,toth05}, from which the $W$ state is
just a one-excitation particular case. In Sec.~\ref{deterministic},
we consider the family of entangled states that could be generated
by means of purely unitary global operations and, in
Sec.~\ref{postselected}, the ones that could be generated by using
postselection.

\section{Collective maps}
\label{collmaps}

Let us consider $N$ ions in a linear Paul trap, cooled down to their
collective motional ground state. We will not concentrate on a specific experimental setup~\cite{leibfried03}, and our derivations could be applied to any ion-trap device. The free-energy Hamiltonian,
$H_0$, describing the $N$ two-level ions and their motion around
their equilibrium positions is
\begin{eqnarray}
H_0 =  \frac{\hbar \omega_0}{2} \sum_{n=1}^N
 \sigma_z^n + \hbar \sum_{j=1}^{N}
\nu_j a^\dagger_j a_j .
\end{eqnarray}
Here, $\sigma^n_z$ are $z$-components of Pauli spin vectors
describing the two-levels with energy gap $\omega_0$, while $a_j$
and $a_j^\dagger$ are the annihilation and creation operators for
the normal modes with frequency $\nu_j$. The interaction between the
internal degrees of freedom of each ion and a collective motional
mode can be induced by laser light of frequency $\omega$,
yielding~\cite{wineland98}
\begin{eqnarray}
H^{n}_{\rm int}= \hbar \lambda_n \sigma_x^{n} \cos (k x_n-\omega t +
\phi_n) . \label{hint}
\end{eqnarray}
Here, $\lambda_n$ is the coupling strength between the laser and the
$n$-th ion, $\sigma^n_x$ are $x$-components of Pauli vectors, $k$ is
the laser wave vector, $x_n$ is the displacement operator with
respect to the equilibrium position, and $\phi_n$ is the phase of
the laser at the location of the $n$-th ion.

We will study the case of homogeneous laser excitation, $\lambda_n =
\lambda$, $\forall n$, and of near resonant coupling, $\omega
\approx \omega_0$. For the sake of simplicity, we will also consider
all $\phi_n = 0$, although this may play an important role when
making experimental considerations. In this case, the Hamiltonian in
the interaction picture, after a rotating-wave-approximation RWA) with
respect to the two internal levels, reads~\cite{wineland98,deng05}
\begin{eqnarray}
H^{\rm I} & = & \frac{\hbar \lambda}{2} \sum_n(\sigma^n_+e^{-i\delta
t}\exp(i k\sum_j b_{nj} \sqrt{\frac{\hbar}{2m\nu_j}}(a_j^\dagger
e^{i\nu_j t}+\nn\\&&a_j e^{-i\nu_j t})) + {\rm H.c.} ,
\end{eqnarray}
where $b_{nj}$ denote the amplitudes of the $j$-th normal mode of
the ion chain in the position expansions, $\delta=\omega-\omega_0$,
and $m$ is the ion mass. In the Lamb-Dicke limit, where all
Lamb-Dicke parameters $\eta_j = k \sqrt{\frac{\hbar}{2 m\nu_j}}$ are
small, the exponential can be expanded and set for a
RWA with respect to the phonon field. In
that case, when the laser frequency is tuned to a particular
collective motional sideband frequency, $\omega=\omega_0 \pm \nu_j$,
we obtain blue and red sideband transition Hamiltonians
\begin{eqnarray}
H^j_{\rm blue} & = & \hbar \tilde{\lambda}_j \sum_n  b_{nj}
(\sigma^n_+ a_j^\dagger + \sigma^n_- a_j)   , \label{falseblue}   \\
H^j_{\rm red} & = & \hbar \tilde{\lambda}_j \sum_n b_{nj}(
\sigma^n_+ a_j + \sigma^n_- a_j^\dagger) , \label{falsered}
\end{eqnarray}
where $\tilde{\lambda}_j = \eta_j \lambda / 2$. The interaction of
Eq.~(\ref{falsered}) appears naturally in the context of CQED, where
a bunch of atoms interact inhomogeneously with a cavity mode and the counter-rotating terms are neglected in the RWA. The dynamics in Eq.~(\ref{falseblue}) is not usual in CQED but can be easily engineered in trapped ions. Only when $\nu_j$ corresponds to
the center-of-mass (COM) mode frequency $\nu_1$, we have $b_{n1} =
b_1$, and we can define the collective angular momentum operators
$L_+ = \sum_n \sigma_+^n$ and $L_- = \sum_n \sigma_-^n$. In that
case, we could rewrite the Hamiltonians of Eqs.~(\ref{falseblue})
and (\ref{falsered}) as
\begin{eqnarray}
H^1_{\rm blue} & = & \hbar \bar{\lambda}_1 (L_+ a^\dagger + L_- a)  \label{blue} \\
H^1_{\rm red} & = & \hbar \bar{\lambda}_1 (L_+ a + L_- a^\dagger)
\label{red},
\end{eqnarray}
where $\bar{\lambda}_1 = b_1 \tilde{\lambda}_1$. The dynamics
associated with the Hamiltonian of Eq.~(\ref{red}) is named after
Tavis and Cummings~\cite{tavis68}, who developed the first
analytical solutions for this model. When we consider a motional
mode different from the center-of-mass one, we could always define
$L^j_+ = \sum_n b_{nj} \sigma_+^n$ and $L^j_- = \sum_n b_{nj}
\sigma_-^n$, but these operators do not satisfy the usual angular
momentum algebra. If we define $L_z\equiv\sum_n \sigma_z^n$ and
$L^2(j)\equiv L_z^2 + \frac{1}{2}(L_+^jL_-^j + L_-^jL_+^j) $, with
$j=0,1,\cdots N-1$, we get
\begin{eqnarray}
& \lbrack L_z, L_{\pm}^j \rbrack = \pm L_\pm^j & , \nonumber \\
& \lbrack L_z, L^2 (j) \rbrack = 0 & ,
\nonumber \\
& \lbrack L_\pm^j, L^2(k) \rbrack \neq 0 & , \,\,  \lbrace j, k
\rbrace \neq 0 .
\end{eqnarray}
In fact, $L_\pm^j$ can still be used to lower and raise the quantum
numbers of $L_z$, but they do not commute with $L^2(j)$. For the
case of the center-of-mass mode, where all commutations relations
are satisfied, we shall denote the eigenstates of $L^2(1)$ and $L_z$
by $|l,m\rangle$, with $l = N/2, N/2-1 ...$, $l > 0$, and $-l\le m
\le l$. States $|l,m\rangle$ are known as the Dicke
states~\cite{dicke54,stockton03,toth05}.

\section{Invariant subspaces}
\label{invsub}

Hamiltonian $H^j_{\rm red}$ conserves the total number of spin and
phonon excitations, and commutes with the excitation number operator
$\hat R\equiv \sum_m a_m^\dagger a_m + L_z + N/2$, while Hamiltonian
$H^j_{\rm blue}$ conserves the difference between the spin and
phonon excitations, hence, it commutes with $\hat B \equiv \sum
a_m^\dagger a_m - L_z + N/2$. It is therefore possible to consider
vibronic subspaces with a fixed number of excitations associated
with $\hat R$ or $\hat B$. If we concentrate on the case $H=H_{\rm
red}^j$, we have the eigenstates $|r,\alpha\rangle$ of $\hat R$,
where $r=0, 1, 2 \cdots$, and $\alpha$ denotes other degeneracy
lifting quantum numbers. We then obtain the block diagonal structure
$H^j_{\rm red} = \oplus_{r=0}^{r=\infty} H_{\rm red}^j(r)$. The
dynamical evolution that is generated by $H^j_{\rm red}$ leaves the
subspaces invariant.

We proceed to discuss certain examples of such invariant subspaces,
for example, the one associated with the case $j=1$. The smallest
eigenvalue of $\hat R$, $r=0$, corresponds to the state $ {\cal
H}_{r=0}=\lbrace |l = N/2,m = -N/2 \rangle |0 \rangle \rbrace$,
i.e., all atoms in their ground state and no phonons in the system.
For the case $r=1$, we have
\begin{eqnarray}
{\cal H}_{r=1} = {\cal H}_{l=N/2}\oplus {\cal H}_{l=N/2-1} ,
\end{eqnarray}
where
\begin{eqnarray}
\!\!\!\!\! {\cal H}_{l=N/2} = \lbrace \ |N/2,-N/2\rangle|1\rangle,\
\ |N/2,-N/2+1\rangle|0\rangle \ \rbrace
\end{eqnarray}
and
\begin{eqnarray}
\!\!\!\!\! {\cal H}_{l=N/2-1} = \lbrace \ |N/2-1,-N/2+1,
\alpha=1\rangle|0\rangle ,..,\nn\\ \ |N/2-1,-N/2+1,
\alpha=N-1\rangle|0\rangle \rbrace .
\end{eqnarray}
The quantum number $\alpha=1, ... , N-1$, lifts the $(N-1)$-fold
degeneracy of the states with $l = -N/2 + 1$. Hence, values of
$\alpha$ enumerate the different angular momentum multiplets. It is
important to stress that $H_{\rm red}^1$ does not mix the different
multiplets and, since $L_- |N/2-1,-N/2+1,\alpha\rangle =0$, there
are no further transitions. This does not follow merely from the
conservation of $\hat R$, which does not forbid transition between
the state $|N/2-1, -N/2+1,\alpha \rangle | 0 \rangle$, which has
terms with one excited atom, and a state with one excited phonon.
This non-mixing property of the multiplets reflects the effect of
quantum {\em irradiance}~\cite{dicke54,devoe96,aharonov01}. The
construction of higher r-number subspaces is straightforward. For
instance, for $r=2$ we shall have ${\cal H}_{r=2} = {\cal
H}_{l=N/2}\oplus{\cal H}_{l=N/2-1}\oplus{\cal H}_{l=N/2-2}$, etc.

A key point in the present work is the use of subspaces which are
bidimensional. In this simple case, the evolution of the system
resembles that of the well known Rabi oscillations. For example, let
us consider the $r=1$ invariant subspace  ${\cal H}_{r=1,l=N/2}$. We
can start with the non-entangled state containing one phonon and
with all the internal spins in their ground states. When we turn on
the Hamiltonian $H_{\rm red}^1$ we obtain an oscillation between the
states
\begin{eqnarray}
\ket{N/2,-N/2}\ket 1 \leftrightarrow \ket{N/2,-N/2+1}\ket 0 .
\label{ssd2}
\end{eqnarray}
State $\ket{N/2 , -N/2+1}$ is a symmetric combination of $N$ terms,
$( |\up\down\down\cdots\rangle + |\down\up\down\cdots \rangle +
\cdots |\cdots\down\down\up \rangle ) / \sqrt{N}$, known as the W
state. Similarly, we could make use of the invariant space ${\cal
H}_{r=1,l=N/2-1,\alpha}$ and, in that case, we would have the
following oscillation
\begin{eqnarray}
\ket{N/2,-N/2+1}\ket 1 \leftrightarrow \ket{N/2,-N/2+2}\ket 0
.\label{ssd3}
\end{eqnarray}
In the general case,  the invariant subspaces can be of higher
dimension, for instance if we start with $n$ phonons in the
multiplet $l=N/2$, the relevant states for $r=n$ becomes, up to
rotations induced by $H_{\rm red}$,
\begin{eqnarray}
&& \!\!\!\!\!\!\!\!\! \ket{-N/2}\ket l \leftrightarrow \ket{-N/2 +
1}\ket {l - 1} \leftrightarrow ... \leftrightarrow \ket{-N/2 +
l}\ket {0} . \nonumber
\\ &&\label{sub}
\end{eqnarray}

So far, we have discussed invariant subspaces which are connected
with the Dicke states and the collective angular momentum operators
with $j=1$. By tuning the laser to couple other motional collective
modes, we can access other $j$ subspaces. As we discuss in the next
section, it is sometimes helpful to combine several steps, and in
each step to couple a different phonon normal mode. For instance, we
can start with the state that contains two different phonon
excitations
\begin{eqnarray}
\ket{\down\down...\down}\ket{1}_i\ket{1}_j,
\end{eqnarray}
couple first the internal levels with the phonon in mode $i$ and
later with the phonon in mode $j$. This process connects us with the
state $L_+^jL_+^i|l=N/2,m=-N/2\rangle$. It is useful to see that in
this type of transitions we have
\begin{eqnarray}
\!\! L_-^j&L_+^i&\ket{N/2,-N/2}=L_-^j\sum_n
b_{ni}\vert\down\down...\down\underbrace{\up}_{n}
\down...\down\rangle\nn\\
&=& \left(\sum_n
b_{ni}b_{nj}\right)\ket{\down\down...\down}=n\delta_{ij}
\ket{\down\down...\down}, \label{lplm}
\end{eqnarray}
where in the last step we used the orthogonality of the normal
modes.

\section{Deterministic Creation of Entangled states}
\label{deterministic}

With the use of $H_{\rm red}$, many relevant states can be created.
We start with the state
\begin{eqnarray}
\ket{N/2,-N/2}\ket 1 _{\nu_1}=\ket{\down\down\down ... \down}\ket 1
_{\nu_1}, \label{state}
\end{eqnarray}
where the $j$-th mode is occupied by a single phonon, and the
internal state is not entangled. A W-state $\ket{W^{N}_1}\equiv
|N/2, -N/2+1\rangle$ can be created by applying a single collective
$\pi / 2$-pulse on the state of Eq.~(\ref{state}). This can be
easily understood by recalling that the above initial state belongs
to the bidimensional Hilbert space, ${\cal H}_{r=1,l=N/2}=\lbrace
|N/2,-N/2\rangle |1\rangle, |W^N_1\rangle |0\rangle \rbrace$. In
principle, by precise control of the duration and intensity of the
laser pulse, a $W$ state can be created between a large number of
ions. In fact, a $W$-state shared by eight ions has been created
recently using a multi-step sequential procedure based on individual
ionic addressing~\cite{haeffner05}. In the present proposal, we
would require the previous preparation of a single phonon in the COM
mode and the application of a single homogeneous global laser pulse.
A related scheme in the context of quantum dots was discussed
recently by Taylor {\it et al.}~\cite{taylor03}.

It is also possible to generate deterministically higher-excitation
Dicke states using other bidimensional invariant subspaces. The
$r=2$ subspace  ${\cal H}_{r=2,l=N/2-1,\alpha}$ is a two-dimensional
space that is spanned by the states
$|N/2-1,-N/2+1\rangle|1\rangle\sim W_1^N$ and
$|N/2-1,-N/2+2\rangle|0\rangle\sim W_2^N$. The first state above is
equivalent, up to local operations, to  the
$W^N_1=|N/2,-N/2+1\rangle$, while the second state, contains terms
with two excited atoms and is equivalent, up to local
transformations, to the second Dicke state
$|N/2,-N/2+2\rangle=W_2^N$. The construction of $W_2^N$ can
therefore proceed as follows. We first obtain as described above
$W_1^N$, using a single pulse. In the second step, we transform
$W_1^N \rightarrow |N/2-1,-N/2+1\rangle$ by changing locally the
phases of each ion. This step requires local addressing implementing
local rotations.  In the final step, we add a single phonon and
apply again $H_{\rm red}^1$ to obtain $W_2^N$, the second member of
the subspace $r=2$. Unfortunately, it seems that for
higher-excitation Dicke states, e.g. $W^N_3$, this ``climbing the
ladder'' method requires also some interaction between the qubits.
To overcome this difficulty we shall discuss other methods.

We consider next extended examples of coupling to other modes and
show that they can be used for generating {\it irradiant states}
~\cite{dicke54,devoe96,aharonov01}. We start with the state
$\psi_0=|N/2,-N/2\rangle|1\rangle_{\nu_j}$ involving one phonon in
the $j$-th mode and all the internal levels in their ground state,
then, we apply the Hamiltonian $H_{\rm red}^j$. The conservation of
$\hat R$ restricts the possible evolution to the subspace of states
with $r=1$, i.e., to the states $L_+^j \psi_0 $ and
$L_-^kL_+^j\psi_0$, with $k=1,2,...N$. However, we notice from
Eq.~(\ref{lplm}) that only terms with $k=j$ do not vanish, hence,
the evolution leads to Rabi oscillations in the bidimensional
Hilbert space $\lbrace \psi_0, L_+^j\psi_0\rbrace$. In this way, we
can generate the family of entangled irradiant states of the form
\begin{eqnarray}
L_+^j\ket{\down\down...\down}, \ \ j=1,2,\cdots N .
\end{eqnarray}
Irradiant states are states that do not emit photons and are thus
more robust to decoherence than radiant states. In our case, this
property is due to the relation in Eq.~(\ref{lplm}). Since the
coupling to the electromagnetic field is through the $L_{\pm}$
operators, as it is for the phonon field, the resultant states are
irradiant \cite{dicke54}. For the the case of two spins, the
resulting state is the EPR state. The experimental feasibility of
irradiance and superradiance in ion traps was discussed and
demonstrated by De Voe and Brewer~\cite{devoe96}.

Having produced certain irradiant states, we can use them as a
starting point for the deterministic generation of an additional
class of states. Irradiant states introduce other bidimensional
invariant subspaces. Since $L_-\ket{\psi_{irr}}=0$, the subspace
$\{\ket{\psi_{irr}}\ket 1,L_+ \ket{\psi_{irr}}\ket 0\}$ is an
invariant subspace of Hamiltonian $H_{\rm red}$ and, therefore, the
second state can be produced by Rabi flipping. This is a new kind of
entangled state which is a superposition of states with two spins in
the upper state,
\begin{eqnarray}
\Psi&\propto&\left(\vert\underbrace{\up\down...\down\up}_{odd}
\down\down...\down\rangle+\text{perm}\right)\nn\\
&-&\left(\vert\underbrace{\down\up...\down\up}_{even}
\down\down...\down\rangle+\text{perm}\right) .
\end{eqnarray}
For the case of four spins the outcome of this process is a GHZ
state. First, we apply $H^3_{\rm red}$, which couples the internal
states with the higher collective mode, $j=N-1=3$, and create the
irradiant state
\begin{eqnarray} L_+^3\ket{\down \down ... \down}=\ket {\up
\down\down\down} -\ket { \down\up\down\down}+\ket {
\down\down\up\down}-\ket { \down\down\down\up} .
\end{eqnarray}
In the next stage, we apply the $j=0$ red Hamiltonian and get
\begin{eqnarray}
L_+^0 L_+^3\ket{\down\down ... \down}= \ket {\up \down\up\down}
-\ket { \down\up\down\up} ,
\end{eqnarray}
which is, up to a local operation, a GHZ state.

\section{Creation of entangled states with Postselection}
\label{postselected}

In the previous section, we have discussed deterministic schemes for
producing irradiant states as well as the lowest Dicke states
(including the $W$ state). However, the full family of Dicke states
could not be generated using only collective unitary
transformations. In the present section, we present another approach
which is useful for producing the full set of Dicke states
\begin{eqnarray}
W_k^N&=&\binom{N}{k}^{-1/2}\left(\vert\underbrace{\up\up..\up}_{k}
\down\down...\down\rangle+\text{perm}\right)\nn\\&\equiv&
\ket{N/2,-N/2+k} .
\end{eqnarray}
The properties of the Dicke states may be of considerable interest
in quantum information, and have been discussed recently by
different authors~\cite{stockton03}, \cite{korbicz05},
\cite{toth05}, \cite{korbicz06}, \cite{wang06}. It can be shown
that the von Neumann entanglement entropy, with respect to a
bi-partite split of $N$ qubits in a Dicke state, increases with
$k$ and saturates gradually for large $k$ values. The behavior of
the (mixed state) entanglement between two qubits \cite{korbicz06}
can be evaluated by considering, for example, the negativity which
increases, almost linearly with $k$.

The basic idea behind our approach is that while a collective
unitary transformation cannot be used to create any Dicke state, a
suitable choice of the initial phonon state can bring us very
close to our goal. In this scheme, however, there will be always a
small error due to mixing with other states. Therefore, unlike the
previous examples, we propose to postselect the phonon state in
order to be certain that the desired Dicke state was produced.

In order to create the Dicke state $W^N_k$, we begin by preparing
the initial state $|N/2,-N/2\rangle|k\rangle_{\nu_0}$. We then apply
the time evolution of the Hamiltonian $H^j_{\rm red}$, which takes
this state into the $r=k$ invariant subspace. It turns out that at a
certain time the probability distribution will be sharply peaked
around a state with zero phonons and $W^N_k$ for the internal
levels. By measuring the number of the phonons it is then possible
to remove the admixture of $W_k^N$ with other states. A procedure to
create and measure the number operator in an ion trap was introduced
by different authors~\cite{cirac93,eschner95,solano05}.
Experimentally, motional Fock states were already produced in the
lab~\cite{leibfried96}, although those techniques required a series
of consecutive Rabi flips.

\begin{figure}
\begin{center}
\includegraphics[width=0.5\textwidth,height=0.4\textheight]{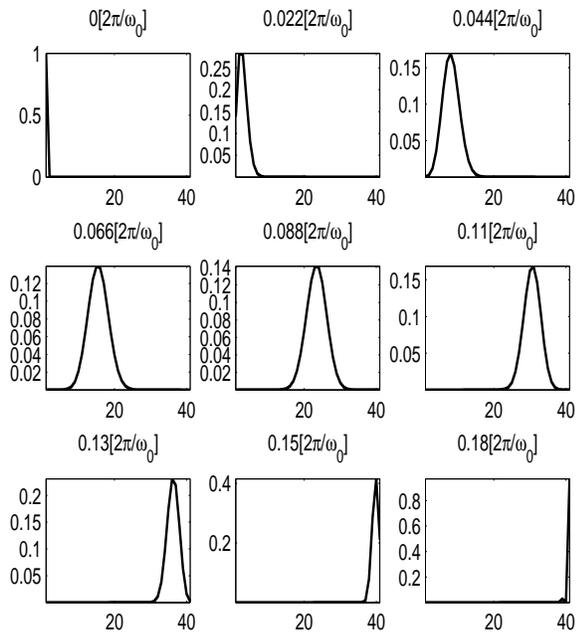}
\end{center}
\caption{Population of various states in Hilbert space as a
function of time for 100 spins and 40 phonons. It can be seen that the population of the last
state is maximal.}\label{pop}
\end{figure}

\begin{figure}
\begin{center}
\includegraphics[width=0.5\textwidth,height=0.12\textheight]{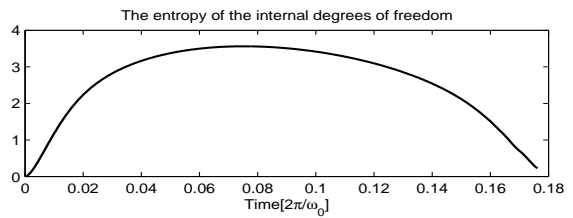}
\end{center}
\caption{The entropy of the internal degrees of freedom as a
function of time for
100 spins and 40 phonons. It can be seen that the final state is
nearly pure (see Fig. 1).}\label{entropy}
\end{figure}

The crucial ingredient in our proposed mechanism is that the purity
of the state, prior to postselection, is high. The fact that the
state containing zero phonons in Eq. (10) is produced with high
probability is shown numerically below. The intuition behind this is
that there is an analogy between these subspaces and the angular
momentum subspaces of $L^2$, though the commutation relations are
different. The $L_x$ operator rotates the spin about the $x$ axis,
producing states with $L_z=\pm l$ with a probability one. In order
to model this dynamics an analogy could be made between this
dynamics in Hilbert space and the dynamics of a particle travelling
between sites with different coupling strength. Since the couplings
are higher at the middle and lower at the edges, the probabilities
are maximal at the edges, see Figs.~\ref{pop} and~\ref{entropy}. We
therefore expect the Hamiltonian $H_{\rm red}$ to rotate the state
between the first and last state in Eq.~(\ref{sub}) with a
probability close to unity.

\begin{figure}
\begin{center}
\includegraphics[width=0.5\textwidth,height=0.2\textheight]{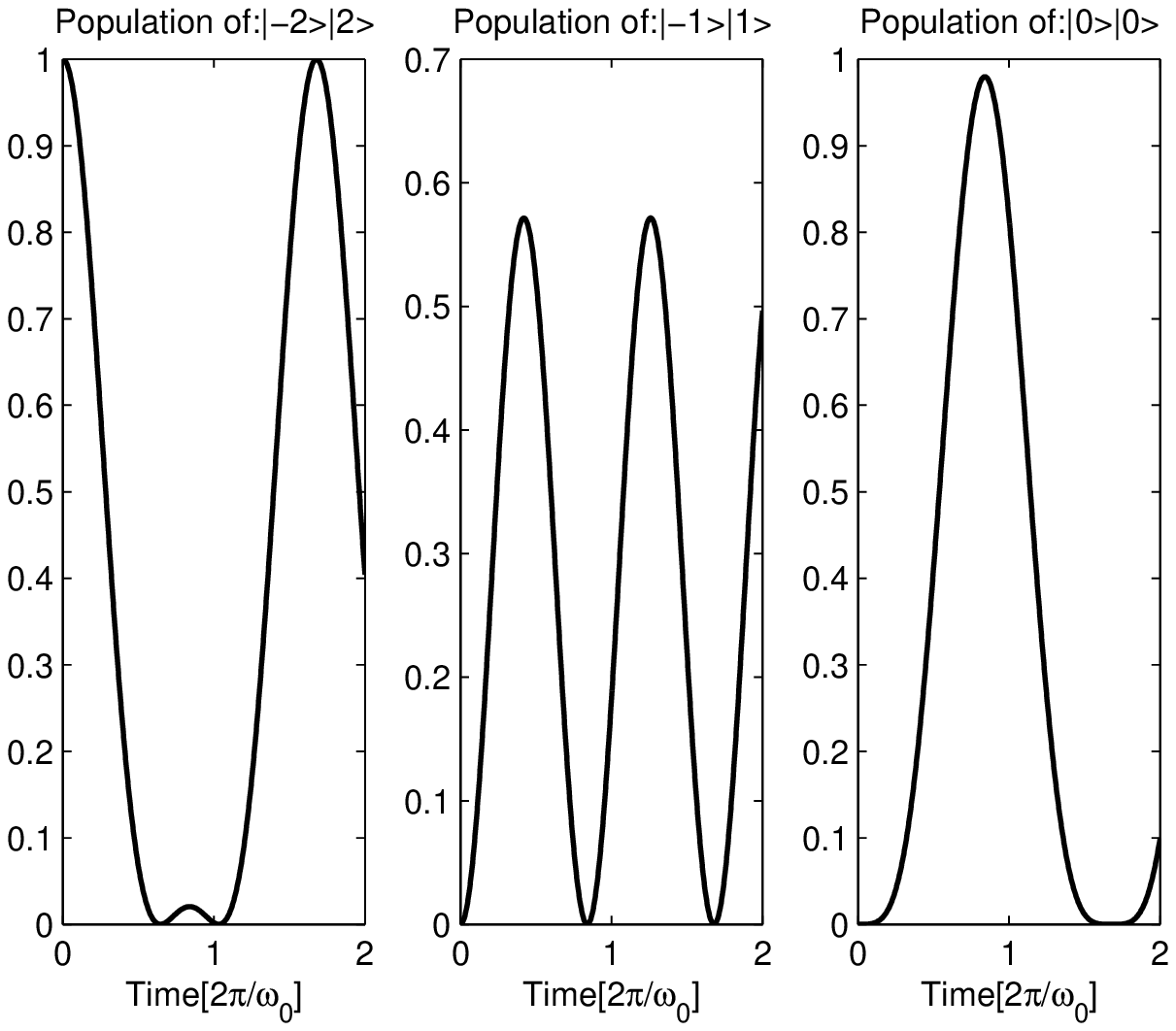}
\end{center}
\vspace*{0cm}
\caption{The probability of states $\ket {-2} \ket 2$, $\ket{-1} \ket{1}$,  $\ket 0 \ket 0$ as a function of time.}\label{pop22}
\end{figure}

\begin{figure}
\begin{center}
\includegraphics[width=0.5\textwidth,height=0.3\textheight]{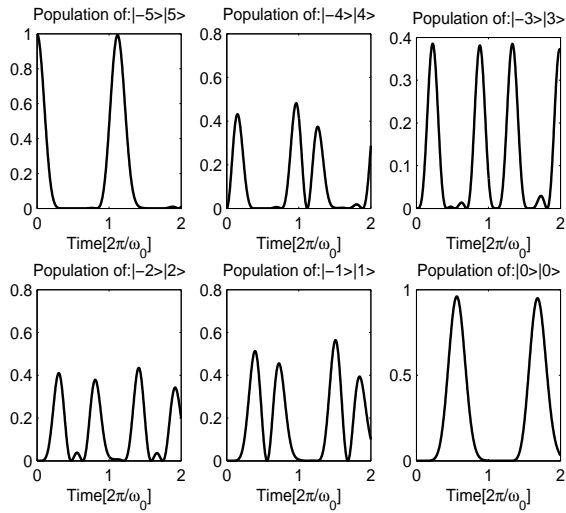}
\end{center}
\caption{The population of the various terms. It can be seen that
except for the first state, only the last state approaches a value
close to $1$. The first state is a state with 10 spins down and 5
phonons}\label{pop55}
\end{figure}

We make now some further considerations concerning our scheme based
on postselection. The only states which are created with high
probability are $\ket {l_z=m}$, where $m$ is the number of phonons.
This is due to the fact that the last state is created with high
probability, therefore, the number of phonons in the first state
determines the final state. The population of states in the subspace
of Eq.~(\ref{sub}) starting with $\ket{-2}\ket{2}$ are shown in
Fig.~\ref{pop22}, where we observe that for specific times the
desired state is obtained with high probability. A similar thing is
observed in Fig. \ref{pop55} starting with $\ket{-5}\ket{5}$. In
spite of the fact that the number of excitations is not negligible
compared to the number of spins, the purity of the final state is
considerably high. This observation may prove very useful for
generating $W_k^N$ states.

In order to increase the purity of the final state the number of
phonons has to be measured and the vacuum state postselected. To
achieve that goal, we consider a recently proposed
technique~\cite{solano05} for sorting a desired motional Fock state
$| N \rangle$ out of any motional distribution. This technique is
based on a suitably designed vibronic scheme in a single ion,
allowing for a restricted dynamics inside a chosen selected JC
subspace $\{ | g \rangle | N + 1 \rangle, | e \rangle | N \rangle
\}$. To adapt it to our present work, we would need an additional
idle ion inside the chain, coupled to the motional mode of interest
and specifically assigned to postselection purposes. Together with
the additional necessity of individual ion addressing for the sake
of manipulation and measurement, these requirements for the idle ion
are at reach by the state-of-the-art present technology in trapped
ions~\cite{haeffner05}.
The proposed scheme described hitherto can also be applied to create
motional number states via the Hamiltonian $H_{\rm blue}$, which
will rotate the state in the proper subspace. Postselecting the spin
state will yield the Fock state state $| N \rangle$ and the number
of spins measured up would indicate the number of motional
excitations $N$.

\section{Conclusions}

In conclusion, we have presented methods of producing entangled
states using homogenous global laser coupling in trapped ion
systems. We have considered two schemes, one based on purely
(deterministic) unitary operations and the other one based on an
ulterior (probabilistic) postselection. Both schemes use the fact
that the Tavis-Cummings model, in its blue- and red-excitation
versions, possesses invariant subspaces. In the deterministic case,
the global laser pulses produce the desired entangled states after
rotations in the associated bidimensional invariant subspaces. In
the probabilistic case, the allowed rotations produce edge states
that are very close to the desired entangled states, requiring a
highly efficient postselection technique. We believe that all
proposed schemes are realistic and at within reach using present
state-of-the-art technology in trapped ions.

\begin{acknowledgements}

We would like to thank Y. Aharonov, H.
Haeffner, I. Klich, B. Groisman, S. Markovitz, S. Nussinov, and M.
Plenio for helpful discussions. Special thanks to J. Eisert for many
useful comments. This work has been supported by the European
Commission under the Integrated Project Qubit Applications (QAP)
funded by the IST directorate as Contract Number 015848. E.S.
acknowledges financial support from DFG SFB 631, EU RESQ and EuroSQIP projects.

\end{acknowledgements}

\end{document}